# Enhanced magnetization in the multiferroic nanocomposites $Bi_{0.9}Gd_{0.1}Fe_{0.9}Mn_{0.05}X_{0.05}O_3$ (X= Cr, Co) thin films.

Mohammed Hadouchi, Françoise Le Marrec, Zakaria Mahhouti, Jamal Belhadi, Mimoun El Marssi and Abdelilah Lahmar*

Laboratoire de Physique de la Matière Condensée, 33 Rue saint-Leu, 80033 Amiens, France

Abstract

In this work we elaborated and characterized nanocomposite thin films using a preselected multiferroic matrix $Bi_{0.9}Gd_{0.1}Fe_{0.9}Mn_{0.1}O_3$ (BGFMO) in the aim to enhance their basic ferroelectric and magnetic properties.

The local ferroelectric character of these nanocomposites thin films was investigated by piezoresponse force microscopy (PFM). Switching behavior for ferroelectric domains was probed by locally manipulating domains. The magnetic investigation showed that the incorporation of transition metal elements in the BGFMO matrix brings out interesting increase in the macroscopic magnetisation at room temperature. Moreover, a spin glass type behavior is observed in $M(T)$ curves for the studied thin films.

Keywords:

multiferroics; $BiFeO_3$-based material; nanocomposite; PFM; magnetic properties

*Corresponding author: abdel.ilah.lahmar@u-picardie.fr

# 1. Introduction

Last decade has seen an increasing interest in magnetoelectric phenomenon, highlighted by the increase number of publications dealing with this topic, either on theoretical or experimental aspects [1,2]. Controlling polarization (respectively magnetization) by magnetic field and/or electric field, known as magnetoelectric effect, allows new range of application such new electronic memory devices, switches, magnetic field sensors or actuators etc.[3,4].

The ME effect can be expressed according to the following equations: [5]

$\Delta P = \alpha \Delta H$ or $\Delta E = \alpha E \Delta H$ [direct ME effect],

$\Delta M = \alpha \Delta E$ [converse ME effect],

where, ($P$) is the polarization, ($H$) applied magnetic field, external electric field ($E$), ($M$) induced magnetization and $\alpha(\alpha E)$ is the ME voltage coefficient.

Noting that to achieve an acceptable $\alpha(\alpha E)$ coefficient, both high ferroelectric and magnetic properties are required. This new challenge, underlines the race towards getting materials with desired properties to meet the new technologies needs.

It should be notice that a number of single phase multiferroic materials has been explored [6–8] but most of them exhibit transition temperatures far below room temperature, thus making potential applications more difficult to realize. Exceptionally, $BiFeO_3$ (BFO), stands out because it has both high ferroelectric- paraelectric transition temperature around 1103 K and high antiferromagnetic to paramagnetic Néel temperature around 643 K [9], promising for a room temperature coupling, that is targeted by industrial application for novel device generation. In this way, BFO has been intensively investigated [10]. The literature examination showed that the BFO bulk form suffers from high leakage currents, attributed to specific structural defects [11,12]. Alternatively, $BiFeO_3$ thin films have seen an ever increasing interest because they exhibit improved ferroelectric as well as magnetic properties over those of bulk materials [10]. Nevertheless, films are still sensitive to the preparation method, which explain the widespread of the reported properties. Despite that monolithic $BiFeO_3$ is claimed to be multiferroic at room temperature, the weak magnetization and moderate polarization prevent from measuring magnetoelectric coupling.

One of the possibilities to enhance ME coupling is the synthesis of artificial nanocomposite multiferoic films materials. Such type of architectures could be achieved by three different ways: (i) multilayer of piezoelectric and magnetostrictive materials; (ii) hierarchical nanostructed composites; or (iii) piezoelectric matrix embedded with magnetic materials. The advantages and drawbacks associate to each case are summarized in ref. [2].

For the way (iii), that concerns the subject of the present study, only few works reported nanocomposites films prepared by a sol-gel process and spin-coating technique.

Obviously, the real problems known for this type of nanocomposites are the low resistivity in the magnetic phase and the difficulty to apply high electric field to induced polarization changes [2]. To avoid this leakage problem, well-defined microstructure is required. Wan et al [13], reported local aggregation or phase separation of the PZT and $CoFe_2O_4$ phase in the nanocomposite films, with both good magnetic and electric properties, as well as magnetoelectric effect.

In the present work, emphasis is placed on using as based matrix for nanocomposites thin films elaboration, the $Bi_{0.9}Gd_{0.1}Fe_{0.9}Mn_{0.1}O_3$ (BGFMO) composition, highlighting its ferroelectric and magnetic properties. The incorporation of $Cr^{3+}$ or $Co^{3+}$ [$Bi_{0.9}Gd_{0.1}Fe_{0.9}Cr_{0.05}M_{0.05}O_3$ (BGFMCr) and $Bi_{0.9}Gd_{0.1}Fe_{0.9}Co_{0.05}M_{0.05}O_3$ (BGFMCo) respectively] has shown to influence microstructure and multiferroic properties of BGFMO.

## 2. Experimental Section

Thin films of the studied materials were elaborated using spin-coating technique on a commercial (111)-Pt/Ti/SiO$_2$/Si substrate heterostructure. The starting reagents were Bi-acetates, Gd acetates, Mn,- Fe, -Co and Cr acetylacetonate, They were dissolved in 2-methoxyethanol to yield a sol concentration of 0.22mol/l. Pyrolysis was conducted on the hot plate at 260 °C. Final annealing was performed in a preheated tube furnace in a saturated oxygen atmosphere at 650 °C for 60 minutes. The phase analysis was characterized at room temperature using a four circle high resolution D8 Discover Bruker diffractometer equipped with a Göbel mirror (Cu $K\alpha$ radiation, $\lambda$=1.5418 Å), by Raman spectroscopy (X-Y spectrometer with Ar$^+$ laser excitation at 514.5 nm, Dilor/Horiba Jobin Yvon, France). The surface morphology and local electromechanical behavior were investigated using a commercial setup NTEGRA Aura (NT-MDT) microscope working in contact mode. For the studies of nanoscale piezo-ferroelectric properties, the microscope is equipped with an external lock-in amplifier (SR-830A Standford Research and Pt/Ir coated tips with nominal stiffness of 7.4 N/m were used (Nanoandmore tips, reference PPP-NCSTPt). The bias voltage is applied the tip whereas the bottom electrode is grounded. The magnetic properties were measured using a commercial Physical Property Measurement System (PPMS DynaCool, Quantum Design). Room temperature hysteresis loops of the investigate samples were carried at H= 1kOe. The temperature dependencies of magnetization *M(T)* were recorded in the temperature interval 2 K <*T*< 300 K and with an applied magnetic field up to *H* = 1kOe in both ZFC (zero-field-cooled) and FC (field cooled) regimes.

## 3. Results and Discussion

*3.1. Structural investigation: X-ray diffraction and Raman spectroscopy studies*

Figure1 shows the XRD patterns of the investigated films. The mother phase (BGFMO) consists of a pure perovskite, and is free from parasitic phases. All observed peaks are indexed in a pseudocubic unit cell. However, for BGFMCo, all reflections of the mother phase are observed with an additional one observed around 30.2°. A similar peak position is observed in the spinel ferrites system with formula $Co_{1-x}Mn_xFe_2O_4$, corresponding to (220) reflexion [14]. Concerning BGFMCr films, the profile of the X-ray diffractogram is different, since only two reflections are observed with the most pronounced one is situated at the vicinity of 29.9°. The observation of similar peak has been reported in Cr doped BFO thin films and was attributed to $Bi_{25}CrO_{40}$ phase [15]. Authors reported that this secondary phase dominate and tends to be the main phase as the concentration of Cr increase. We can evoke herein some similarity with this work.

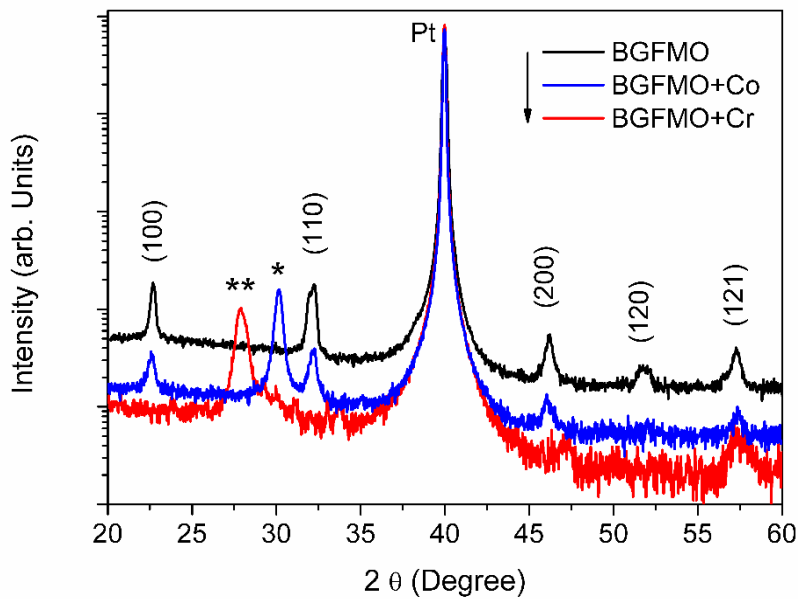

*Figure.1: XRD pattern of BGFMO, BGFMCo , andBGFMCr deposited on (111)-Pt/Ti/SiO$_2$/Si, Pt denotes peaks belonging to the substrate and (\*) , (\*\*) additional peaks.*

In order to get more structural informations which are not accessible by XRD measurements, we have performed Raman spectroscopy investigation. The obtained spectra are compared in Fig.2.

As it can be shown in the figure, the mother phase BGFMO presents two large bands situated at around 620 cm$^{-1}$ and in the range from 450 to 550 cm$^{-1}$. We have assigned these features to respectively symmetric and antisymmetric stretching of the basal oxygen ions of the $Mn^{3+}O_6$ octahedra associated with the Jahn–Teller distortion. More discussion about this behavior and the effect of the presence of manganese on the BFO host lattice can be found in our previous works [16,17]. For BGFMO with Cobalt, the spectrum is similar to the mother phase's one, albeit a shoulder is observed around 676 cm$^{-1}$. A close look to the few previous studies concerning Raman of cobalt oxides, reveal that Choi et al have already reported a strong band at 672 cm$^{-1}$ in CoO [18]. This mode exists also in $CoFe_2O_4$ at ~ 690 cm$^{-1}$ and can shift depending on the incorporation of other transition metal and the related composition. Such mode was assigned to the $A_{1g}(1)$ mode reflecting the stretching vibration of $Fe^{3+}$ and $O^{2-}$ ions in octahedral sites in the nanoparticles spinel [19].

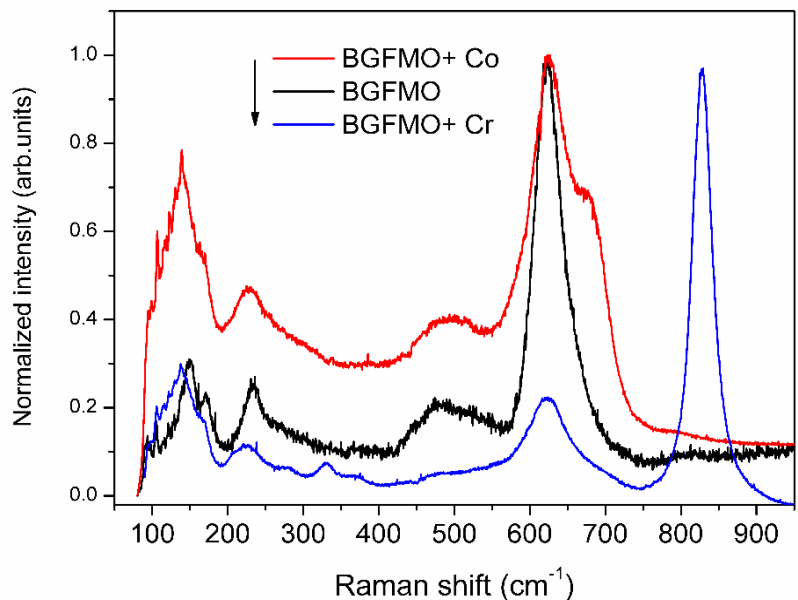

*Figure 2: Raman spectra of the investigated specimens*

Otherwise, the Raman spectra of Cr-phase reveals two remarkable changes. Firstly, the feature developed around 622 cm$^{-1}$ shows a considerable reduction of its intensity compared to the mother phase. The reduction of this intensity let imply a reduction of $Mn^{3+}$ ions, favoring the presence of the non-JT active $Mn^{4+}$ ions. Secondly, the spectrum of Cr-phase exhibits a new band developed at Ca. 827 cm$^{-1}$ which may come from Cr–O band [15].

## 3.2. AFM and PFM microstructure analysis

Figure 3(a) shows the surface morphologies for the three different compositions obtained by atomic force microscopy. All films were crack-free. A fine and rounded grained structure with a smooth surface topography is observed for all specimens and a root mean square (RMS) roughness for a 10 × 10 μm$^2$ scan area is found to be between 5 and 6 nm. Noting that the microstructure is affected by the incorporation of metal transition element. As it can be seen from the images, the microstructure of the mother phase is uniform. For the film with $Co^{3+}$ grains seem to be get together to form clusters separated by vacant spaces. The microstructure for the film with $Cr^{3+}$ is completely different. The incorporation of the $Cr^{3+}$ in BFO matrix revealed the existence of foreign phase agglomerated as dunes and well distributed over the matrix.

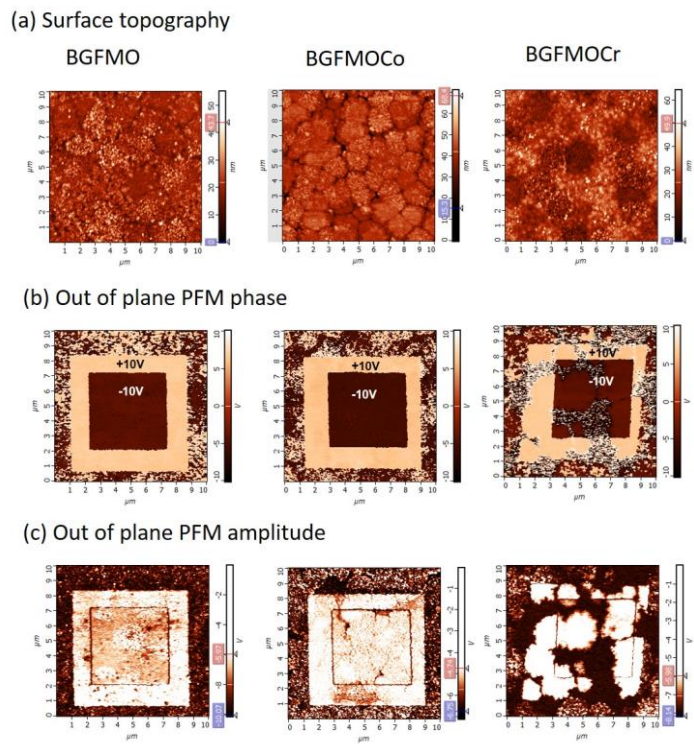

*Figure 3. (a) AFM surface topography micrographs of BGFMO, BGFMCo, and BGFMCr. Out of plane PFM phase(b) and amplitude(c) images after poling experiments. The 7.5×7.5 μm$^2$ and 5×5 μm$^2$ squares were, respectively, positively and negatively poled (±10V).*

The local piezoelectric properties of the films were investigated using piezoelectric force microscopy (PFM). In order to get valuable information about the effect of $Co^{3+}$ and $Cr^{3+}$ incorporation on piezoelectric behavior of BGFMO compound, we first study and compare out-of-plane PFM images for the three samples. Figure 4 shows typical PFM images obtained on the three as-grown films. For the mother phase and the film with $Co^{3+}$, PFM measurements reveal a clear piezoelectric contrast associated with the direction of the polarization, with bright (pale brown) and dark (dark brown) contrasts corresponding to domains having polarization components along [00-1] and [001] directions, respectively. This indicates that films are not-pre-poled as usually encounter for epitaxial films. Concerning the film with $Cr^{3+}$, besides observing a 180° phase contrast, PFM image reveals a third region with a black / white contrast (indicative of absent of signal) which is correlated to the topography. We will discuss the presence of the third region later.

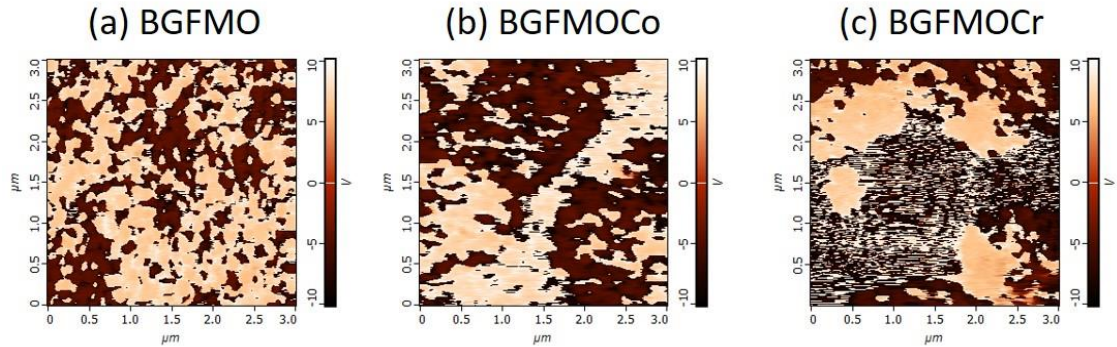

*Figure 4 : Out of plane PFM phase images recorded on the as-grown thin films (before poling experiments).*

In order to investigated localized polarization switching behaviors for the three films, a DC bias voltage was applied to the conductive PFM-tip probe. A 7.5×7.5μm² square were initially written by applying a positive bias voltage, followed by the writing of an internal 5×5μm² square negatively poled. The result of these poling experiments are shown in figure 3b and 3c. As highlighted by the well-defined squares observed after poling experiments (see figure 3b), polarization is successfully performed for the mother phase BGFMO and the film with $Co^{3+}$. These results clearly evidence poled domains corresponding to downward and upward polarization stable states and demonstrate that the polarization process is completely reversible. These results corroborate the macroscopic polarization switching observed for the mother composition [20]. The ferroelectric character of BGFMO and BGFMO-Co films is confirmed by PFM amplitude signal as displayed in figure 3c. Images exhibit typical features: bright regions which correspond to piezoelectric vibration of the manipulated domains whereas dark regions (that is no vibration amplitude) which are attributed to domain walls between areas of opposite polarizations. Concerning BGFMO-Cr film, the squares observed after poling experiments are not completely well-defined. The PFM amplitude image reveals clearly piezoelectric inactive regions (dark regions) which can be correlated to the topography. This observation is further corroborated by the local piezoelectric loop acquired on the three films (not shown here). Well-saturated loops are observed for BGFMO and BGFMO-Co films, whatever is placed the AFM tip over the film surface, which is not the case for BGFMO-Cr film for which the presence of saturated loops is topographic dependent. For this latter film, the coexistence of ferroelectric and non-ferroelectric regions corroborates XRD and Raman investigations.

*3.4. Magnetic investigations*

*3.4.1. Room temperature magnetic investigation*

The magnetic hysteresis curves at room temperature for the studied specimens are shown in Figure 5. A well-developed S-shaped *M-H* curve was obtained for BGFMO, in comparison to pure BFO, with quite higher saturation magnetization (Ms~20 emu.cm$^{-3}$) [21]. In a previous work [20] we have reported that presence of $Gd^{3+}$ was indirectly responsible for the enhancement of magnetization comparing de BFO thin films and this magnetic improvement seems to be related to the change of tilting angle between $Fe^{3+}$ caused by the change of the nature of octahedral distortion. Herein, as it can be seen form figure 5, the incorporation of $Co^{3+}$ or $Cr^{3+}$ improved considerably the saturation magnetization which reached 80 emu.cm$^{-3}$ . However, the presence of Cobalt seems not affect so much the remnant magnetization. In contrast, the presence of $Cr^{3+}$ enhances both magnetization and coercive field (Mr~15emu.cm$^{-3}$ ; Hc~ 0.3kOe).

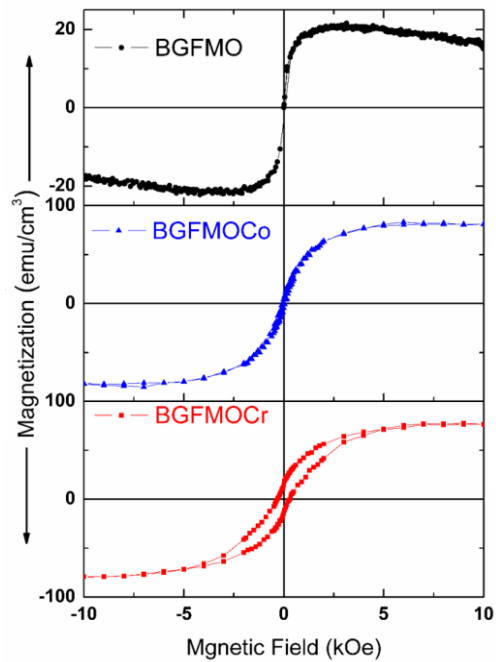

*Figure 5: Room temperature M-H curves at measured for BGFMO, BGFMCo, and BGFMCr thin films*

We shall now discuss the origins of magnetization improvement observed in the prepared nanocomposite films. Recall that Goodenough et al. [22] reported that the substitution of Mn ions by other transition metal such Co, Cr or Ni induced ferromagnetism in $LaMnO_3$. Concerning the BGFMCr films, a clear ferromagnetic behavior is depicted at RT. On considering BFGM as mother phase, we expected that $Mn^{3+}(3d^4)$ ions are partially substituted by isoelectronic $Cr^{3+}$ ($3d^3$) which have the same electronic configuration as $Mn^{4+}$ ($e_g$, $t^3_{2g}$). In the system $LaMn_{1-x}Cr_xO_3$, Zhang et al [23] invoked the possibility of double exchange interaction (DEI) $Mn^{3+}$- O- $Cr^{3+}$, contrarily to the work of Gundakaram et al. [24] on $LnMn_{1-x}Cr_xO_3$(Ln= La, Pr, Nd, Gd) where no DEI could be expected. The literature survey showed controversial reports about the contribution of $Cr^{3+}$ in DEI in the manganites [25–27]. However, the partial substitution of $Fe^{3+}$ by $Cr^{3+}$ could induce the 180° Cation-anion-cation super-exchange interactions between $Fe^{3+}$-O-$Cr^{3+}$. This type of magnetic interaction is ferromagnetic according to Goodenough - Kanamori - Anderson (GKA) semi-empirical rules [28–30] which might explain the open hysteresis loop observed for BGFMCr thin films. For instance, XRD investigation and AFM showed the presence of two phases probably BGFMO and $Bi_{25}CrO_{40}$. Further, Raman investigation showed the reduction of JT distortion with a predominant Cr-O vibration mode. The assumption of the presence of $Co^{2+}$ and $Mn^{4+}$ is not excluded. In such case, the formation of $Mn^{4+}$ could be compensated either by the formation of $Fe^{2+}$, Bi vacancies or annihilation of oxygen vacancies. The presence of all these possibilities does not allow us to give an objective interpretation of the origin of the ferromagnetic behavior observed in BGFMCr thin films.

Regarding the nanocomposite BGFMCo films, a narrow hysteresis loop is obtained with Mr~5emu.cm$^{-3}$. Such behavoir could be considered as a ferrimagnetic like (or soft ferromagnetic) one. A similar ferrimagnetic *M–H* loop was observed in Co doped BFO [31] or in double-perovskite $Bi_2(FeMn)O_6$ thin films [32,33]. It is worthwhile to note the nanosized $CoFe_2O_4$ are reported to exhibit similar magnetic behavior at RT [34]. Noting that the simultaneous presence of $Mn^{3+}$, $Co^{3+}$ and $Fe^{3+}$ is not always a sufficient condition to improve magnetization at room temperature. Recently we have shown that in the BFO-$LaMn_{0.5}Co_{0.5}O_3$ system, the simultaneously presence of $Mn^{3+}$ and $Co^{3+}$ in the Fe-sites doesn't improve the magnetization of BFO at room temperature [35], because the result structural edifice do not permit the magnetic interactions to be active. In addition, the Raman investigation undertaken for BGFMCo sample highlight the existence of JT distortion induced by the presence of $Mn^{3+}$ which is still active similarly to the mother matrix. Thus, the presence of $Mn^{4+}$ is excluded, otherwise JT distortion should be reduced. So, it is legitimate to infer that magnetization improvement observed in BGFMCo nanocoposite films can be imputed to only presence of spinel secondary phase.

*3.4.2. Magnetic behavior as function of the temperature*

For further investigating the magnetic properties of the prepared samples, we have performed temperature dependence of magnetization in Zero-Field-cooled (ZFC) and Field Cooled (FC) modes and *M-H* curves at different temperatures as shown in Figs. 6 and 7 respectively. For BGFMO, it is clearly seen that the ZFC and FC curves present bifurcation at 100 Oe starting from a temperature of T = 274 K. This behavior is typical of spin-glass type [36,37]. Noting that similar behavior is observed for BGFMCr and BGFMCo thin films at the same field, but with different bifurcation temperatures, i.e., 205 and 265 K for BGFMCo and BGFMCr respectively. The plots show that for these films, the divergence between ZFC and FC curves decreases with increasing the field. Simultaneously, the bifurcation temperature moves to lower temperature until disappearance at 4 kOe for BGFMO and BGFMCr, but it persists for BGFMCo. Such overall behavior is linked to the glassiness characteristic [38,39]. Similar magnetic behavior was reported for pure BFO thin films [37]. However, the existence of bifurcation in *M(T)* curves only for BGFMCo thin film at 4 kOe (not observed for the two other films), lets surmise the presence of a secondary magnetic phase, as depicted previously in X-ray diffraction and Raman investigations. We believe that the existence of spinel $CoFe_2O_4$ is the most probable candidate. In fact, similar bifurcation in the *M-T* curves even at 5 kOe was observed in the nanosized $CoFe_2O_4$ and was attributed to superparamagnetic and ferrimagnetic behavior of the particles [34]. Additionally, the observed blocking temperature around 150 K for BGFMCo thin film at 100 Oe deserves attention. Indeed, similar blocking temperature was observed in spherical $CoFe_2O_4$ nanocrystals where the magnetic properties are strongly related to the nanocrystals shapes [40]. At very low temperature, i.e., T ≈7 K, ≈10 K, and ≈15 K, for BGFMO, BGFMCo, and BGFMCr respectively, we observe an abrupt jump of magnetization especially at 500 Oe and higher fields, which is explained by the occurrence of week ferromagnetic component. Similar magnetization jumps were reported for BFO single crystal [36] and thin films [37]. This weak ferromagnetism is confirmed in the *M-H* curves for BGFMO and BGFMCo by the presence of an open hysteresis at low temperature. While for BGFMCr, the loop became larger ( see Fig. 6).

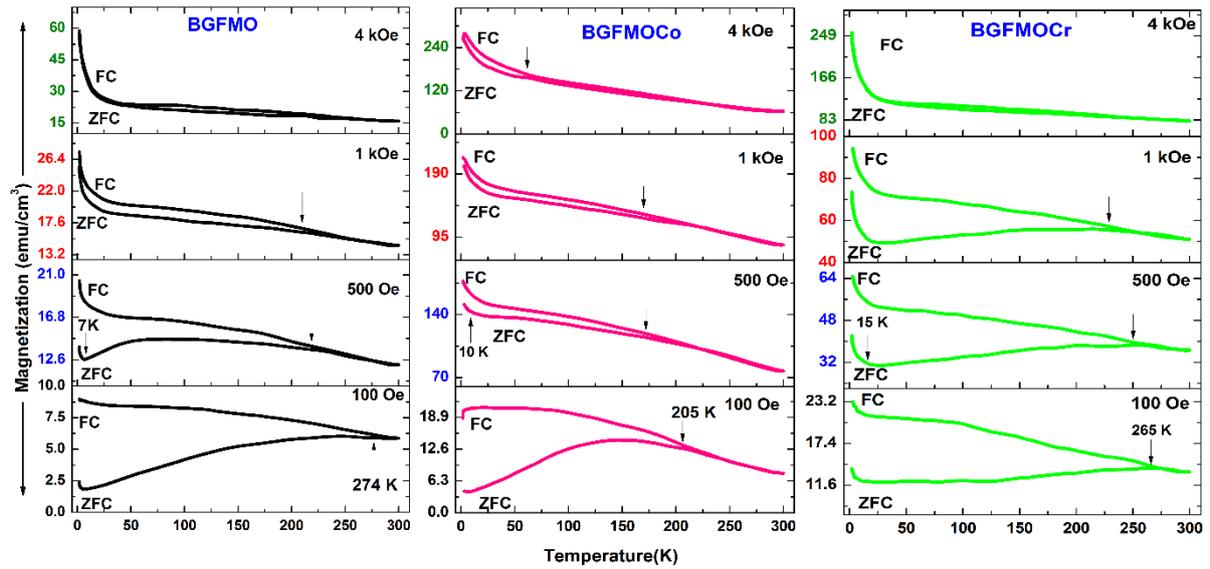

Fig. 6. Temperature dependence of magnetization for BGFMO, BGFMCo and BGFMCr thin films

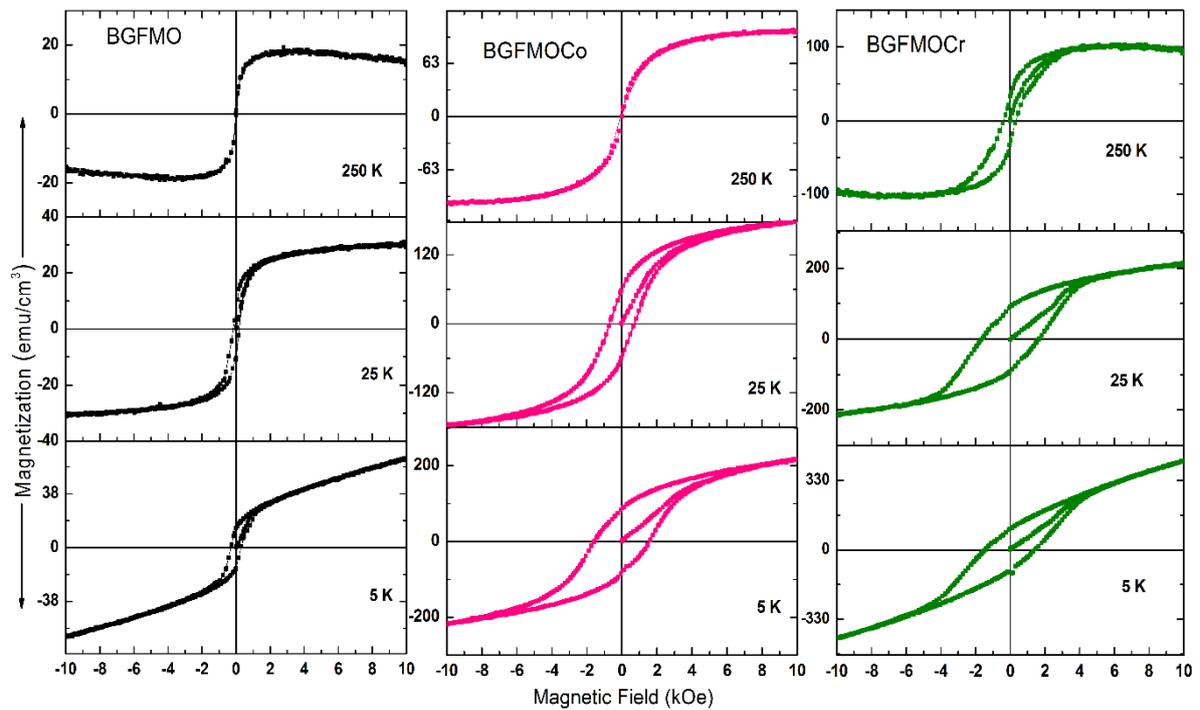

Fig. 7. *M-H* curves at different temperatures for BGFMO, BGFMCo, and BGFMCr thin films

## 4. Conclusion

In the present work, the morphology, structure, and multiferroic properties of Co and Cr doped $Bi_{0.9}Gd_{0.1}Fe_{0.9}Mn_{0.1}O_3$ composite thin films were investigated. We have shown that the incorporation of Co in the BGFMO matrix induced a change in the microstructure with uniformed topography instead of the appearance of a secondary phase for Cr inclusion. Nice local ferroelectric properties are obtained for BGFMCo at room temperature similarly to the mother phase, while a non-ferroelectric secondary area was detected in the case of BGFMCr nanocoposite films.

Both nanocomposite films exhibit improved RT- magnetization compared to the initial matrix. However, the presence of cobalt didn't seems to affect considerably the remnant magnetization. Whereas, the phase with Cr exhibits improved both magnetization and coercive field.

To get more information about the effect of Co and Cr on the magnetic properties, temperature dependence of magnetization in ZFC and FC modes and *M-H* loops at different temperatures were performed on the investigated samples. The obtained results of *M(T)* curves show the occurrence of spin-glass-type behavior in all thin films. However, these results lest suggested that the improvement of magnetization as well as the change in the magnetic behavior are related directly to the nature of the secondary phase.

Based on these findings, we conclude that these nanocomposites bear promising for extended functionalities.

## Acknowledgments

This work was supported by the European H2020-MC-RISE-ENIGMA action (N°778072).

## References

[1] C.A. Vaz, Electric field control of magnetism in multiferroic heterostructures, J. Phys. Condens. Matter. 24 (2012) 333201.
[2] R.C. Kambale, D.-Y. Jeong, J. Ryu, Current status of magnetoelectric composite thin/thick films, Adv. Condens. Matter Phys. 2012 (2012).
[3] J. Wang, J. Neaton, H. Zheng, V. Nagarajan, S. Ogale, B. Liu, D. Viehland, V. Vaithyanathan, D. Schlom, U. Waghmare, Epitaxial $BiFeO_3$ multiferroic thin film heterostructures, Science. 299 (2003) 1719–1722.
[4] H. Zheng, J. Wang, S. Lofland, Z. Ma, L. Mohaddes-Ardabili, T. Zhao, L. Salamanca-Riba, S. Shinde, S. Ogale, F. Bai, Multiferroic $BaTiO_3$-$CoFe_2O_4$ nanostructures, Science. 303 (2004) 661–663.
[5] J. Ma, J. Hu, Z. Li, C. Nan, Recent progress in multiferroic magnetoelectric composites: from bulk to thin films, Adv. Mater. 23 (2011) 1062–1087.
[6] H. Schmid, Introduction to the proceedings of the 2nd international conference on magnetoelectric interaction phenomena in crystals, MEIPIC-2, Ferroelectrics. 161 (1994) 1–28.
[7] M. Bichurin, Short introduction to the proceedings of the 3 RD international conference on magnetoelectric interaction phenomena in crystals, MEIPIC-3, Ferroelectrics. 204 (1997) xvii–xx.

[8]  W. Eerenstein, N. Mathur, J.F. Scott, Multiferroic and magnetoelectric materials, Nature. 442 (2006) 759.

[9]  S. Kiselev, R. Ozerov, G. Zhdanov, Sov. Phys. Dokl., (1963).

[10] R. Ramesh, N.A. Spaldin, Multiferroics: progress and prospects in thin films, in: Nanosci. Technol. Collect. Rev. Nat. J., World Scientific, 2010: pp. 20–28.

[11] W. Eerenstein, F. Morrison, F. Sher, J. Prieto, J. Attfield, J. Scott, N. Mathur, Experimental difficulties and artefacts in multiferroic and magnetoelectric thin films of $BiFeO_3$, $Bi_{0.6}Tb_{0.3}La_{0.1}FeO_3$ and $BiMnO_3$, Philos. Mag. Lett. 87 (2007) 249–257.

[12] Y. Wang, C.-W. Nan, Enhanced ferroelectricity in Ti-doped multiferroic Bi Fe O 3 thin films, Appl. Phys. Lett. 89 (2006) 052903.

[13] J. Wan, X. Wang, Y. Wu, M. Zeng, Y. Wang, H. Jiang, W. Zhou, G. Wang, J.-M. Liu, Magnetoelectric $CoFe_2O_4$–Pb (Zr, Ti)$O_3$ composite thin films derived by a sol-gel process, Appl. Phys. Lett. 86 (2005) 122501.

[14] R. Kambale, P. Shaikh, N. Harale, V. Bilur, Y. Kolekar, C. Bhosale, K. Rajpure, Structural and magnetic properties of $Co_{1−x}Mn_xFe_2O_4$ ($0 \leq x \leq 0.4$) spinel ferrites synthesized by combustion route, J. Alloys Compd. 490 (2010) 568–571.

[15] H. Deng, H. Deng, P. Yang, J. Chu, Effect of Cr doping on the structure, optical and magnetic properties of multiferroic $BiFeO_3$ thin films, J. Mater. Sci. Mater. Electron. 23 (2012) 1215–1218.

[16] A. Lahmar, S. Habouti, M. Dietze, C.-H. Solterbeck, M. Es-Souni, Effects of rare earth manganites on structural, ferroelectric, and magnetic properties of $BiFeO_3$ thin films, Appl. Phys. Lett. 94 (2009) 012903.

[17] A. Lahmar, M. Es-Souni, Sequence of structural transitions in $BiFeO_3$–$RMnO_3$ thin films (R= Rare earth), Ceram. Int. 41 (2015) 5721–5726.

[18] H.C. Choi, Y.M. Jung, I. Noda, S.B. Kim, A study of the mechanism of the electrochemical reaction of lithium with CoO by two-dimensional soft X-ray absorption spectroscopy (2D XAS), 2D Raman, and 2D heterospectral XAS− Raman correlation analysis, J. Phys. Chem. B. 107 (2003) 5806–5811.

[19] N. Sanpo, J. Tharajak, Y. Li, C.C. Berndt, C. Wen, J. Wang, Biocompatibility of transition metal-substituted cobalt ferrite nanoparticles, J. Nanoparticle Res. 16 (2014) 2510.

[20] A. Lahmar, S. Habouti, C. Solterbeck, M. Dietze, M. Es-Souni, Multiferroic properties of Bi 0.9 Gd 0.1 Fe 0.9 Mn 0.1 O 3 thin film, J. Appl. Phys. 107 (2010) 024104.

[21] A. Lahmar, K. Zhao, S. Habouti, M. Dietze, C.-H. Solterbeck, M. Es-Souni, Off-stoichiometry effects on $BiFeO_3$ thin films, Solid State Ion. 202 (2011) 1–5.

[22] J.B. Goodenough, A. Wold, R. Arnott, N. Menyuk, Relationship between crystal symmetry and magnetic properties of ionic compounds containing $Mn^{3+}$, Phys. Rev. 124 (1961) 373.

[23] L. Zhang, G. Feng, H. Liang, B. Cao, Zhu мeihong, and YG Zhao, JMMM. 219 (2000) 236.

[24] R. Gundakaram, A. Arulraj, P. Vanitha, C. Rao, N. Gayathri, A. Raychaudhuri, A. Cheetham, LETTER TO THE EDITOR: Effect of Substitution of $Cr^{3+}$ in Place of $Mn^{3+}$ in Rare-Earth Manganates on the Magnetism and Magnetoresistance: Role of Superexchange Interaction and Lattice Distortion in $LnMn_{1−x}Cr_xO_3$, (1996).

[25] T. Kimura, Y. Tokura, R. Kumai, Y. Okimoto, Y. Tomioka, Relaxor behavior in manganites, J. Appl. Phys. 89 (2001) 6857–6862.

[26] I.O. Troyanchuk, M.V. Bushinsky, V.V. Eremenko, V.A. Sirenko, H. Szymczak, Magnetic phase diagram of the system of manganites $Nd_{0.6}Ca_{0.4}(Mn_{1−x}Cr_x)O_3$, Low Temp. Phys. 28 (2002) 45–48. doi:10.1063/1.1449184.

[27] Y. Sun, X. Xu, Y. Zhang, Effects of Cr doping in $La_{0.67}Ca_{0.33}MnO_3$ : Magnetization, resistivity, and thermopower, Phys. Rev. B. 63 (2000). doi:10.1103/PhysRevB.63.054404.

[28] J.B. Goodenough, Magnetism And The Chemical Bond, John Wiley And Sons, NY-London, 1963.

[29] J. Kanamori, Superexchange interaction and symmetry properties of electron orbitals, J. Phys. Chem. Solids. 10 (1959) 87–98. doi:10.1016/0022-3697(59)90061-7.

[30] P.W. Anderson, New Approach to the Theory of Superexchange Interactions, Phys. Rev. 115 (1959) 2–13. doi:10.1103/PhysRev.115.2.

[31] Y.-S. Chiang, C.-S. Tu, P.-Y. Chen, C.-S. Chen, J. Anthoniappen, Y. Ting, T.-S. Chan, V.H. Schmidt, Magnetic and phonon transitions in B-site Co doped $BiFeO_3$ ceramics, Ceram. Int. 42 (2016) 13104–13112. doi:10.1016/j.ceramint.2016.05.097.

[32] E.-M. Choi, T. Fix, A. Kursumovic, C.J. Kinane, D. Arena, S.-L. Sahonta, Z. Bi, J. Xiong, L. Yan, J.-S. Lee, H. Wang, S. Langridge, Y.-M. Kim, A.Y. Borisevich, I. MacLaren, Q.M. Ramasse, M.G. Blamire, Q. Jia, J.L. MacManus-Driscoll, Room Temperature Ferrimagnetism and Ferroelectricity in Strained, Thin Films of $BiFe_{0.5}Mn_{0.5}O_3$, Adv. Funct. Mater. 24 (2014) 7478–7487. doi:10.1002/adfm.201401464.

[33] Q. Xu, Y. Sheng, M. Khalid, Y. Cao, Y. Wang, X. Qiu, W. Zhang, M. He, S. Wang, S. Zhou, Q. Li, D. Wu, Y. Zhai, W. Liu, P. Wang, Y.B. Xu, J. Du, Magnetic interactions in $BiFe_{0.5}Mn_{0.5}O_3$ films and $BiFeO_3/BiMnO_3$ superlattices, Sci. Rep. 5 (2015) 9093.

[34] V. Masheva, M. Grigorova, N. Valkov, H.J. Blythe, T. Midlarz, V. Blaskov, J. Geshev, M. Mikhov, On the magnetic properties of nanosized $CoFe_2O_4$, J. Magn. Magn. Mater. 196–197 (1999) 128–130. doi:10.1016/S0304-8853(98)00688-X.

[35] A. Lahmar, Multiferroic properties and frequency dependent coercive field in $BiFeO_3$ - $LaMn_{0.5}Co_{0.5}O_3$ thin films, J. Magn. Magn. Mater. 439 (2017) 30–37. doi:10.1016/j.jmmm.2017.04.096.

[36] M.K. Singh, W. Prellier, M.P. Singh, R.S. Katiyar, J.F. Scott, Spin-glass transition in single-crystal $BiFeO_3$, Phys. Rev. B. 77 (2008). doi:10.1103/PhysRevB.77.144403.

[37] C.-J. Cheng, C. Lu, Z. Chen, L. You, L. Chen, J. Wang, T. Wu, Thickness-dependent magnetism and spin-glass behaviors in compressively strained $BiFeO_3$ thin films, Appl. Phys. Lett. 98 (2011) 242502. doi:10.1063/1.3600064.

[38] S.F. Edwards, P.W. Anderson, Theory of spin glasses, J. Phys. F Met. Phys. 5 (1975) 965–974. doi:10.1088/0305-4608/5/5/017.

[39] S.F. Edwards, P.W. Anderson, Theory of spin glasses. II, J. Phys. F Met. Phys. 6 (1976) 1927–1937. doi:10.1088/0305-4608/6/10/022.

[40] Q. Song, Z.J. Zhang, Shape Control and Associated Magnetic Properties of Spinel Cobalt Ferrite Nanocrystals, J. Am. Chem. Soc. 126 (2004) 6164–6168. doi:10.1021/ja049931r.